\documentclass{IOS-Book-Article}
\usepackage{xcolor}
\usepackage{mathptmx}
\usepackage{graphicx}
\usepackage{url}

\def\hb{\hbox to 10.7 cm{}}
\newcommand\rev[1]{#1}

\begin{document}

\pagestyle{headings}
\def\thepage{}

\begin{frontmatter}              

\title{Contagion dynamics of extremist propaganda in social networks}

\markboth{}{May 2017\hb}

\author[A]{\fnms{Emilio} \snm{Ferrara}%
\thanks{Corresponding Author: Emilio Ferrara, USC Information Sciences Institute, 4676 Admiralty Way \#1001, Marina del Rey, CA (USA) 90292; E-mail:
emiliofe@usc.edu}}

\runningauthor{E. Ferrara}
\address[A]{University of Southern California, Information Sciences Institute, USA}

\begin{abstract}
Recent terrorist attacks carried out on behalf of ISIS on American and European soil by lone wolf attackers or sleeper cells remind us of the importance of understanding the dynamics of radicalization mediated by social media communication channels. In this paper, we shed light on the social media activity of a group of twenty-five thousand users whose association with ISIS online radical propaganda has been manually verified. 
By using a computational tool known as \textit{dynamic activity-connectivity maps}, based on network and temporal activity patterns, we investigate the dynamics of social influence within ISIS supporters.  
We finally quantify the effectiveness of ISIS propaganda by determining the adoption of extremist content in the general population and draw a parallel between radical propaganda and epidemics spreading, highlighting that information broadcasters and influential ISIS supporters generate highly-infectious cascades of information contagion.
Our findings will help generate effective countermeasures to combat the group and other forms of online extremism.
\end{abstract}

\end{frontmatter}
\markboth{May 2017\hb}{May 2017\hb}

\section{Introduction}
Researchers in the computational social science community have recently demonstrated the importance of studying online social networks to understand our society~\cite{lazer2009life}.
New powerful technologies are usually harbinger of abuse, and online platforms are no exception: social media have been shown to be systematically abused for nefarious purposes~\cite{ferrara2015manipulation}.
As online social environments yield plenty of incentives and opportunities for unprecedented, even ``creative'' forms of misuse, single individuals as well as organizations and governments have systematically interfered with these platforms, oftentimes driven by some hidden agenda, in a variety of reported cases: 

\begin{itemize}
\item During crises, social media have been effectively used for emergency response; but fear-mongering actions have also triggered mass hysteria and panic~\cite{gupta20131,ferrara2015manipulation}.

\item Political conversation has been  manipulated by means of orchestrated astroturf campaigns~\cite{ratkiewicz2011detecting,metaxas2012social} even during election times~\cite{howard2016bots,bessi2016social}.

\item Anti-vaccination movements~\cite{subrahmanian2016darpa,tangherlini2016mommy}, as well as conspiracy (and other anti-science) theorists~\cite{bessi2015science,del2016spreading}, took social media by the storm and became responsible for a major health crisis in  the United States~\cite{forbes2015vaccine}.

\item Social media bots (non-human automated accounts) have been used to coordinate attacks to successfully manipulate the stock market, causing losses in the billions of dollars~\cite{hwang2012socialbots,ferrara2015manipulation,ferrara2016rise,varol2017online}.

\item Some governments and non-state actors have been active on social media to spread their propaganda. In some cases, they have allegedly ``polluted'' these platforms with content to sway public opinion~\cite{ferrara2015manipulation,dong2017managing,king2017chinese}, or to hinder the ability of social collectives to communicate, coordinate, and mobilize~\cite{serrato2016influence}.
\end{itemize}

Especially related to the last point, researchers have been recently devoting more attention to issues related to online propaganda campaigns~\cite{schiermeier2015terrorism, reardon2015terrorism, allendorfer2015isis}.
Increasing evidence provided by numerous independent studies suggests that social media played a pivotal role in the rise in popularity of the \textit{Islamic State  of Iraq and al-Sham} (viz. ISIS)~\cite{fisher2015jihadist, stern2015isis, weiss2015isis, ferrara2016predicting, ferrara2017the}. 
Determining whether ISIS benefitted from using social media for propaganda and recruitment was central for many research endeavors~\cite{berger2013who,  berger2015isis, magdy2016failedrevolutions}. 

Analyses by Berger and Morgan suggested that a restricted number of highly-active accounts (500-1000 users) is responsible for most of ISIS' visibility on Twitter~\cite{berger2015isis}.
However, Berger's subsequent work suggested that ISIS' reach (at least among English speakers) has stalled for months as of the beginning of 2016, due to more aggressive account suspension policies enacted by Twitter~\cite{berger2016isis}. 
Other researchers tried to unveil the roots of support for ISIS, suggesting that ISIS backers discussed Arab Spring uprisings on Twitter significantly more than users who stood against ISIS~\cite{magdy2016failedrevolutions}.

{\rev These early investigations all share one common methodological limitation, namely that to collect social media data they start from keywords known to be associated to ISIS~\cite{berger2015isis, berger2016isis,  magdy2016failedrevolutions}. 
This strategy has been widely adopted in a previous research aimed at characterizing social movements~\cite{gonzalez2011dynamics, conover2013digital, varol2014evolution}}. However, we argue that it is not sufficient to focus on keyword-based online chatter to pinpoint to relevant actors of radical conversation. 

{\rev In fact, our recent results~\cite{ferrara2017the} suggest that radical propaganda revolves around four independent types of messanging: \textit{(i)} theological and religious topics; \textit{(ii)} violence; \textit{(iii)} sectarian discussion; and, \textit{(iv)} influential actors and events.
Here is a series of examples of possible biases introduced by the keyword-centric approach: 

\begin{itemize}
\item Some studies~\cite{magdy2016failedrevolutions} focused on religion-based keyword lists, but most terms typically associated to religion are not necessarily used in the context of extremism.
\item Other studies~\cite{berger2015isis, berger2016isis} focused on influential actors or events; this can introduce biases due to the focus on popular actors rather than the overall conversation.
\item Further noise can be introduced by tweets that simply link to news articles reporting on events; although these tweets may contain keywords in the predefined watchlist, they clearly do not represent extremist propaganda efforts.
\item Finally, focusing on pre-determined keywords could cause incomplete data collection by missing topics of discussion that can emerge dynamically and do not adopt any of the predefined key terms.
\end{itemize} 

In this work, we will leverage an alternative data collection and curation approach: we will start from a large set of Twitter users that are known to be associated to or symphatizers of ISIS. We will then collect their activity over a large time span of over one year to obtain a complete characterization of their extremist propaganda efforts.
}

\subsection{Contributions of this work}
This study aims to address the two following research questions: 

\begin{itemize}
\item[\textbf{RQ1}:]
Can we define a solid methodological framework and suggest good practices for data collection, validation, and analysis to study online radicalization?.
\end{itemize}

{\rev Setting such best practices will hopefully steer the information sciences and computational social sciences research communities in the direction of producing more rigorous and reproducible work. One contribution of our work is to address this issue by focusing on manually-verified set of ISIS supporter accounts.}

\smallskip
After describing how we collected and curated the dataset object of this study, we move forward to investigate the dynamics of online propaganda adopted by ISIS. Using computational social science tools to gauge online extremism, we aim to answer the following second research question:

\begin{itemize}
\item[\textbf{RQ2}:]
What types of network and temporal patterns of activity reflect the dynamics of social influence within ISIS supporters? And, can we quantify the adoption of extremist content in the general population?.
\end{itemize}

Our findings will help generate effective countermeasures to combat the group and other forms of online extremism.

\section{Data Collection and Curation}
Due to the the limits of keyword-based data collection approaches we highlighted above, in this work we exclusively rely on data and labels obtained by using a procedure of manual data curation and expert verification. 
We obtained a list of Twitter accounts whose activity was labeled as supportive of the Islamic State by the online crowd-sourcing initiative called \emph{Lucky Troll Club}. 
The goal of this initiative was to leverage annotators with expertise in Arabic languages to identify accounts affiliated to ISIS and report them to Twitter to request their suspension. Twitter's anti-abuse team manually verifies all suspension requests, and grants some based on evidence of violation of Twitter's Terms of Service policy against the usage of the platform for extremist purposes. 

We verified that 25,538 accounts present in the \emph{Lucky Troll Club} list have been suspended by Twitter in the period between March 17, 2015 and June  9, 2015. In this study, we focus only on this subset of twenty-five thousand accounts: we consider their activity to be unequivocally linked to the Islamic State, as determined by the two-step manual verification process described above. For each account, we have at our disposal information about the suspension date, as well as the number of followers of that user as of the suspension date. 
{\rev The anonymized data focus of this study can be made available to the interested researchers upon request. For contact information and further details about the dataset please visit: \url{http://www.emilio.ferrara.name/datasets/}}

\subsection{Twitter data collection}\vspace{-3mm}
The next step of our study consisted in collecting data related to the activity of the 25,538 ISIS supporters on Twitter. To this purpose, we leveraged the Observatory on Social Media (OSoMe) data source set up by our collaborators at Indiana University~\cite{davis2016osome}, which continuously collects the Twitter data stream from the \textit{gardenhose} API (roughly a 10\% random sample of the full Twitter data stream). Using this large data stream avoids known issues derived by using the public Twitter stream API which serves only less than 1\% of the overall tweets~\cite{morstatter2013sample}.
	
We obtained all tweets present in the OSoMe database that have been posted by any of the twenty-five thousand ISIS accounts prior to their suspension. We also included all retweets these tweets generated, and all tweets containing mentions to such set of ISIS supporters.
The resulting dataset that we will study consists of 3,395,901 tweets. Almost 1.2 million of these tweets was generated by the ISIS accounts during the period between January 2014 and June 2015. We found that a total of 54,358 distinct other users has retweeted at least once one of the twenty-five thousand ISIS supporters. This amounts for the remainder of about 2.2 million tweets in our dataset.

\smallskip
\noindent Summarizing,  we identified two best practices to answer \textbf{RQ1}:

\begin{itemize}
\item Favor starting from a manually-verified list of users involved in online propaganda, radicalization efforts, or recruitment, rather focusing on keyword-based searches. When such manually validated lists are not available, human annotations can be generated by means of services such as Amazon Mechanical Turk.
\item Use large social media data streams, when available, rather than small samples that can be biased. Services like the Indiana University's OSoMe database~\cite{davis2016osome} can provide data sources especially valuable for Twitter-based social media studies. Alternatively, the Twitter Search API can yield comprehensive data around specific users or topics, provided that the search is limited to short time frames.
\end{itemize}

{\rev 
\subsection{Limitations and strategic choices of this study}
The largest majority of the tweets in our dataset, over 92\%, is in Arabic. This introduces a number of important challenges, from both technical and methodological perspectives.
From a technical standpoint, analyzing Arabic content would require sophisticated natural language processing (NLP) techniques as well as accurate sentiment analysis tools capable of extracting emotional information.
NLP toolkits with support for Arabic content are only recently starting to be developed~\cite{green2010better}, while Arabic-based sentiment analysis is still in its early developmental stage~\cite{korayem2012subjectivity}.

From a methodological point of view, the analysis of Arabic-language material would require interpreters with sufficient domain knowledge in extremism-related issues to yield useful and unbiased insights from the data. Access to such experts is not always available, and thus the need to develop alternative strategies of enquiry emerges.

For these reasons, in the rest of the paper our analysis will rely exclusively upon language-agnostic techniques: in particular, we will focus on statistical properties of information diffusion networks, which were proven very useful in our prior studies on criminal networks~\cite{agreste2016network}, as well as temporal patterns of information diffusion, which we already exploited to study the interplay between ISIS' activity online and offline~\cite{ferrara2017the}. 

Due to the simplifications introduced by our strategy of enquiry, we call for caution in the interpretation of our results. However, we believe that our approach  will help identify important areas of research that warrant further development (such as Arabic-based NLP and sentiment analysis toolkits), as well as yield valuable insight about ISIS social media operations and other forms of online extremist propaganda.

}

\vspace{-3mm}
\section{Results}

\begin{figure}[t] 
\includegraphics[width=\columnwidth]{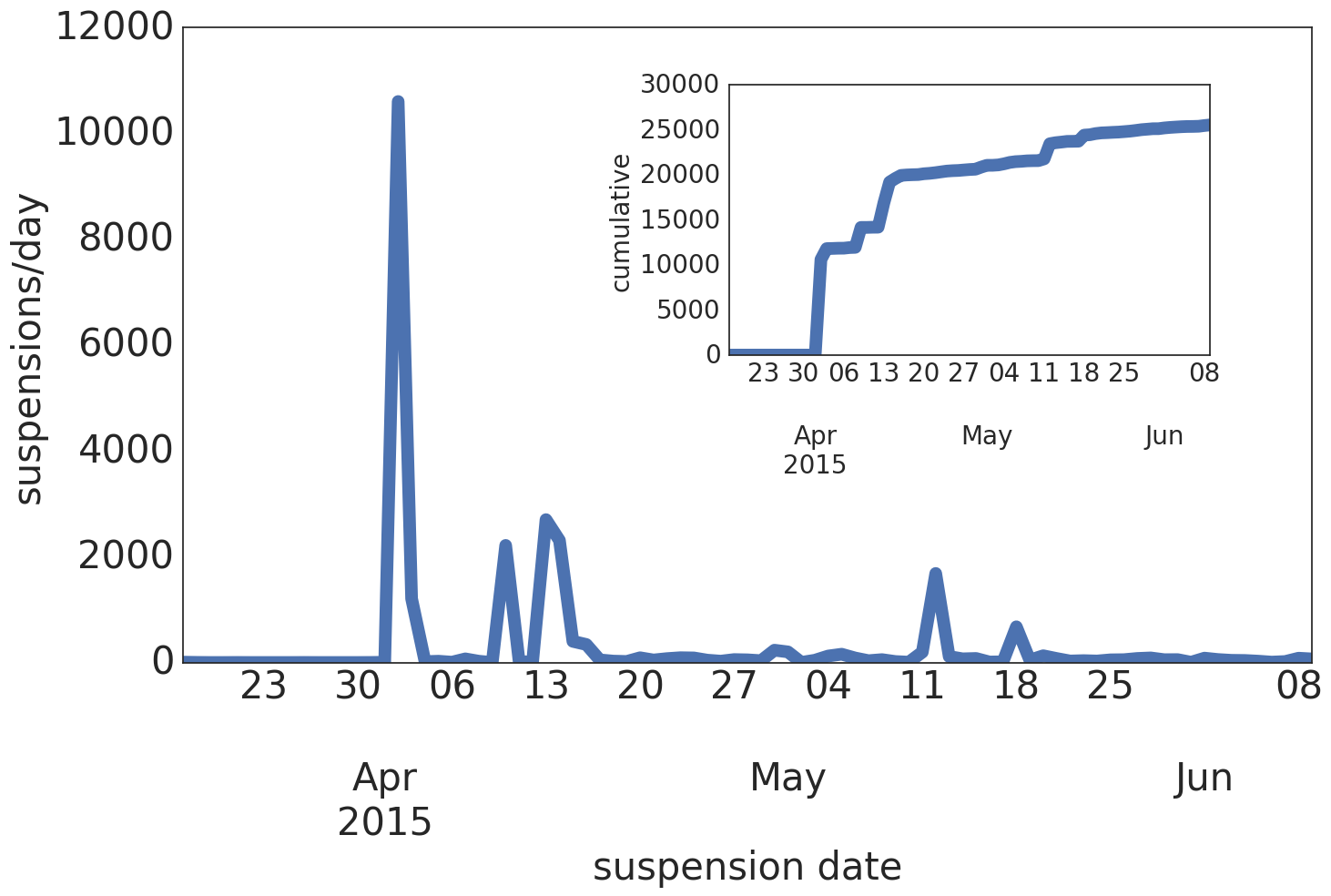}
\caption{Timeline of suspensions per day of the 25,538 ISIS-supporting accounts. The main figure shows the daily suspension counts, while the inset figure shows the cumulative count.}
\label{fig:suspension_timeline}
\end{figure}

We next report our investigation to address \textbf{RQ2}. We study the activity of ISIS supporters and sympathizers on Twitter by means of the data collected as described above.

\subsection{Activity and Support of ISIS Accounts}\vspace{-3mm}

The first form of validation that we performed pertains the mechanisms of suspension of ISIS accounts on Twitter. As we mentioned before, we identified a set of twenty-five thousand accounts related to ISIS that have been suspended: Fig.~\ref{fig:suspension_timeline} shows the timeline of suspensions of the accounts under investigation. The suspension period occurred throughout almost three months, with the first suspensions occurring on March 17, 2015 and the last occurring on June  9, 2015. 
After this date, none among the twenty-five thousand accounts in our list is anymore active on Twitter. 
Account suspensions appear to occur in batches, some more substantial than others, with a significant spike of suspensions (over ten thousands) occurred on April 2, 2015. 
Our findings are consistent with \textit{The Guardian}'s report that, between April 2015 and February 2016, Twitter's anti-abuse task force suspended more than 125,000 accounts linked to ISIS~\cite{guardian2016isis}.

\begin{figure}[t] 
\includegraphics[width=\columnwidth]{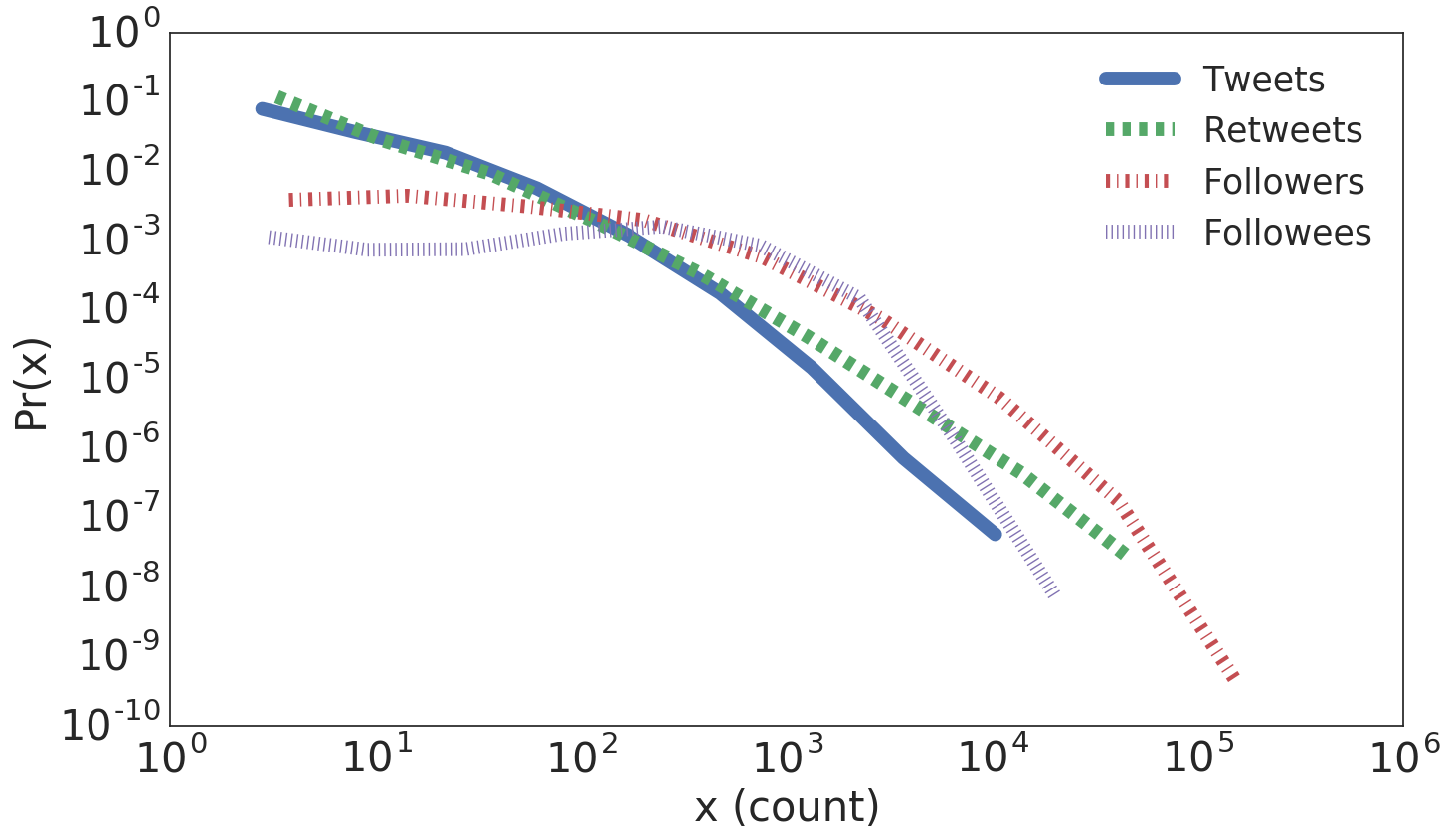}
\caption{\rev Probability distributions of the number of tweets posted by ISIS supporters, generated retweets, and the number of followers and followees at the time of their suspension.}
\label{fig:distributions}
\end{figure}

Recent literature reported contrasting evidence about the activity and popularity of ISIS supporters on Twitter~\cite{berger2015isis, berger2016isis}. As for today, it is unclear whether ISIS accounts obtained a significant support on Twitter, and to what extent they were active. Berger and collaborators first found that ISIS presence was very pervasive on Twitter during 2014-2015~\cite{berger2015isis}, and later suggested that only a core of 500-2000 ISIS users was active after that period~\cite{berger2016isis}. 
To shed light on this question, we calculated the distribution of the number of followers and followees (friends) of ISIS accounts at the time of their suspension, along with the number of tweets and retweets they generated.
{\rev For each of these four variables, we calculated their probability distribution $Pr(x)$. These probability distributions display the probabilities of values taken by the four variables (\textit{i.e.,} tweets, retweets, followers, followees) in our dataset: they can be thought as \textit{normalized frequency distributions}, where all occurrences of outcomes for each distribution sum to 1.
The results are shown in Fig.~\ref{fig:distributions}. }

Let us discuss support and activity  separately.
Concerning ISIS support on Twitter, we notice that the distribution of followers of ISIS accounts exhibits the long tail typical of Twitter~\cite{kwak2010twitter} and other social networks (\textit{cf.} yellow dash-dotted line in Fig.~\ref{fig:distributions}). This skewed distribution has mean $\mu = 516$ ($\sigma=1,727$), median $Q_2=130$, and lower and upper quartiles $Q_1=37$ and $Q_3=401$.
This means that the majority of accounts has a limited number of followers (for example, one quarter of the users has less than 37 followers), yet a significant number of ISIS supporters managed to obtain a large number of followers before getting suspended (in fact, the upper quartile of users has more than 400 followers). The presence of this broad distribution of followership suggests that influence and radicalization operations of Islamic State supporters on Twitter were successful for at least several thousand of their accounts. This is in line with Berger's early results discussed above~\cite{berger2015isis}. Another interesting insight is yielded by the distribution of followees (\textit{cf.} green dotted line in Fig.~\ref{fig:distributions}): differently from the distribution of followers, this distribution shows an unexpected upward trend in the regime between 100 and 1,000 followees. This characteristic behavior has been associated to forms of social network manipulation, for example attempts to create rings or cliques in which multiple accounts under the control of a same entity all follow each other to reciprocally increase their visibility and followership~\cite{ratkiewicz2011detecting}. We thus suggest that ISIS accounts enacted strategies to artificially enhance their visibility by strengthening one another social networks.

\begin{figure}[t] 
\includegraphics[width=\columnwidth]{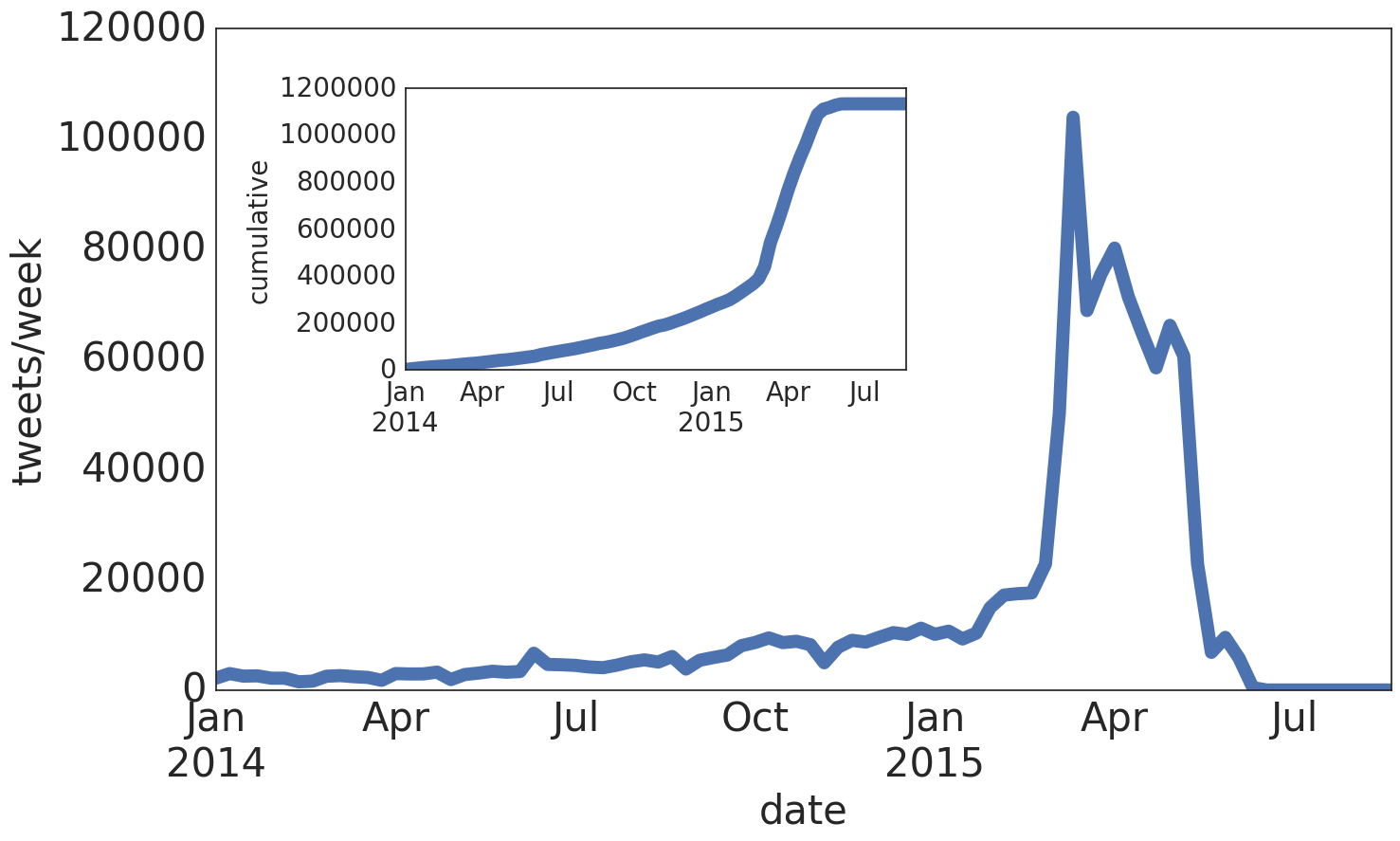}
\caption{Timeline of total number of tweets generated by ISIS supporters. The main figure shows the number of tweets per week; the inset figure shows the cumulative count.}
\label{fig:total_tweets}
\end{figure}

A significant portion of ISIS accounts was very active on Twitter: both distributions of tweets (\textit{cf.} solid blue line) and retweets (\textit{cf.} red dashed line) shown in Fig.~\ref{fig:distributions} exhibit the typical power-law shape common to social networks with heterogeneous activity patterns. This implies that a significant fraction of users posted and retweeted large amounts of tweets. For example, at least 1\% of the ISIS users posted at least 30 tweets during the observation period;  similar figures hold for retweeting. More importantly, there appears to be a strong core constituted by a few dozen accounts who posted and retweeted hundreds of tweets in the same period. This set of very active ISIS supporters we found is compatible with what reported by Berger's first study~\cite{berger2015isis}. Furthermore, there seems to be a handful of accounts with thousands of tweets and retweets, suggesting the likely presence of some social media bot used to enhance the volume of content generated by ISIS and its spreding on Twitter~\cite{ferrara2016rise}.

To investigate ISIS accounts' activity further, we extrapolated the time series of the total volume of tweets and retweets posted every week by the ISIS users under investigation: the result is shown in Fig.~\ref{fig:total_tweets}. It's worth noting that our observation window spans 1.5 years and starts on January 2014 when the firsts among the twenty-five thousand ISIS supporters became active on Twitter: Although the activity volume slowly builds up over time, it is only in early March 2015 (15 months into our observation period) that the volume of tweets per week drastically spikes. 

In the early  regime between January 2014 and March 2015 the volume of tweets associated to  ISIS accounts spans 1,000 to 10,000 per week. This increases nearly tenfold after March 2015, with a spike of over 100,000 tweets per week, and an average of over 60,000 tweets per week in the period between March and May 2015. This period concurs with the period of strongest Twitter suspensions shown in Fig.~\ref{fig:suspension_timeline}, suggesting a timely reaction of Twitter to fight the activity of ISIS users on the platform. Indeed, the volume of tweets per week produced by these accounts drops in early June, and goes to zero, as expected, in the late period of observation when fewer and fewer of the ISIS accounts under investigation are left unchecked on Twitter. Cumulatively, the ISIS users under investigation produced almost 1.2 million tweets during the observation period (\textit{cf.} inset of Fig.~\ref{fig:total_tweets}).

Our findings pinpoint to the power of the crowd-sourcing volunteer initiative that set to bring up to Twitter's attention these accounts. However, there is no evidence to quantify how many (if any) of the ISIS supporters not recorded by the \textit{Lucky Troll Club} operation were independently suspended by Twitter anti-abuse team.

\subsection{Dynamic Activity-Connectivity Maps}\vspace{-3mm}

\begin{figure}[t] 
\includegraphics[width=\columnwidth]{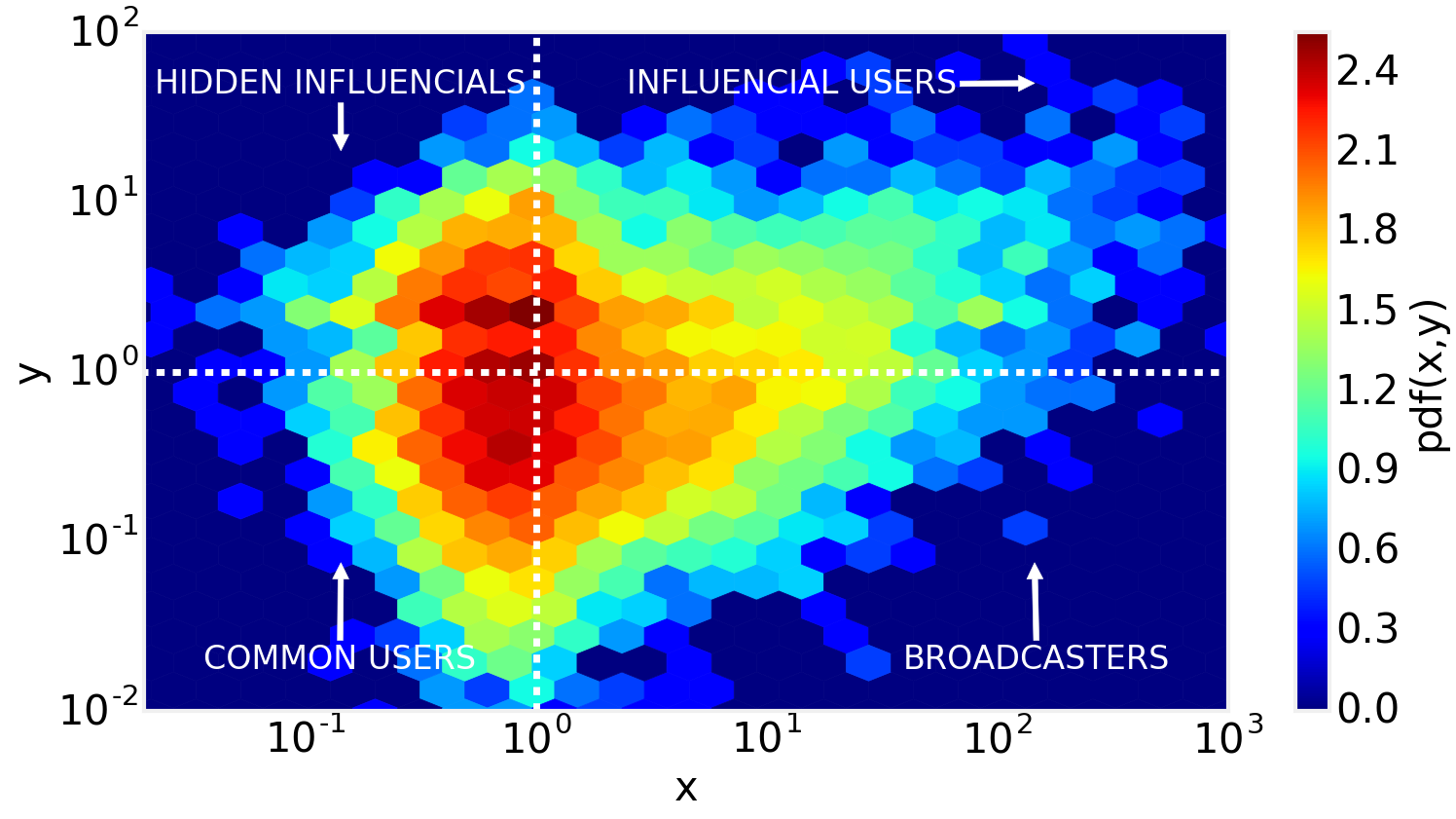}
\caption{\rev The \textit{dynamic activity-connectivity map} shows the Joint Probability Density Function $pdf(x,y)$ of the connectivity growth $x$ and the activity rate $y$ as defined in Equation~\ref{eq:1}.}
\label{fig:conn_activity}
\end{figure}

{\rev We further address \textbf{RQ1} by offering a powerful tool to information scientists and computational social scientists: the \textit{dynamic activity-connectivity (DAC) maps}. DAC maps allow to study influence and authority dynamics of online extremism in social networks with temporal activity patterns.

Fig.~\ref{fig:conn_activity} shows an example of \textit{dynamic activity-connectivity map}. We developed DAC maps as dynamic variants of the map proposed by Gonzalez-Bailon and collaborators---see Figure 4 in \textit{Broadcasters and Hidden Influentials in Online Protest Diffusion}~\cite{gonzalez2013broadcasters}.
The key intuition behind this tool is to allow investigate what effect the progression of activity levels of an user has of their connectivity evolution (and viceversa).

In a DAC map,} for a given user $u$, $x_u$ and $y_u$ are defined as 

\begin{equation}
x_u = \frac{1+\delta f_u}{1+\delta F_u} \qquad  \mbox{and} \qquad y_u= \frac{1+ m_u}{1+ M_u}.
\label{eq:1}
\end{equation}

We use the notation $f_u$ and $F_u$ to identify the number of followers and friends, respectively, of a user $u$. 
The variation of followers and friends of user $u$ over a period of time $t$ are thus defined as  $\delta f_u = \frac{f_u^{max} - f_u^{min}}{t}$ and $\delta F_u = \frac{F_u^{max} - F_u^{min}}{t}$; the length of time $t$ is defined as the number of days of $u$'s activity, measured from registration to suspension (this varies from user to user).
Finally, $m_u$ is the number of mentions user $u$ received by others, and $M_u$ is the number of mentions user $u$ made to others, during $u$'s activity period. 
All values are added to the unit to avoid zero-divisions and allow for logarithmic scaling (i.e., in those cases where the variation is zero). {\rev The third dimension, the``heat'' (the color intensity) in the map, represents the joint probability density $pdf(x,y)$ for users with given values of $x$ and $y$.} The plot also introduce a bin normalization to account for the logarithmic binning. 

{\rev The two dimensions defined over the \textit{dynamic activity-connectivity map} are interpreted as follows: the x-axis represents the growth of connectivity formation, and the y-axis conveys the rate of messaging activity. 
In general, we would expect that in a dynamic activity-connectivity map,} the bulk of the joint probability density mass would be observed in the neighborhood of $(1,1)$, that hosts the majority of accounts for which the variation of the two dimensions is comparable.

{\rev DAC maps are ideal tools to addres our second research question (\textbf{RQ2}): they capture at the same time network and temporal patterns of activity, and they can help understand how connectivity variations affects social influence (and viceversa),  which are dynamics at the core of our investigation.}

Let us discuss the two dimensions of Figure~\ref{fig:conn_activity}, namely \textit{connectivity growth} and \textit{activity rate}, separately.
The \textit{connectivity growth} is captured by the $x$ axis and, in our case, ranges roughly between $10^{-2}$ and $10^3$. 
Users for which $x>1$ (i.e., $10^0$) are those with a followership that grows much faster than the rate at which these users are following others. In other words, they are acquiring social network popularity (followers) at a fast-paced rate. Note that, if a user is acquiring many followers quickly, but s/he is also following many users at a similar rate, the value of $x$ will be near 1. This is a good property of our measure because it is common strategy on social media platforms, especially among bots~\cite{ferrara2016rise,bessi2016social}, to indiscriminately follow others in order to seek for reciprocal followerships. Our dynamic activity-connectivity map will discriminate users with fast-growing followerships, who will appear in the right-hand side of the map, from those who adopt that type of reciprocity-seeking strategy. The former group can be associated with highly popular users with a fast-paced followership growth. 
According to Gonzalez-Bailon and collaborators~\cite{gonzalez2013broadcasters} this category is composed by two groups: \textit{influential users} and \textit{information broadcasters}, depending on their activity rates.
Values of $x<1$ indicate users who follow others at a rate higher than that they are being followed; the fall in the left-hand side of the map.
According to Gonzalez-Bailon and collaborators, these are mostly the \textit{common users}, although the so-called \textit{hidden influentials} also sit in this \textit{low-connectivity} regime.

As for what concerns the $y$ axis, it measures the \textit{activity rate}, i.e.,  the rate at which a user receives mentions versus how frequently s/he mentions others.
Users with values of $y>1$ are those who receive systematically more mentions with respect to how frequently they mention others. This group of users  can be referred to as \textit{influentials},  i.e., those who are referred to significantly more frequently than others in the conversation; they fall in the upper region of the map, and according to Gonzalez-Bailon \textit{et al.}, depending on their connectivity growth can be divide in influential ($x>1$) and hidden influential ($x<1$) users.
Conversely, users with values of $y<1$ are those who generate increasingly more mentions over time, either because they reply to many tweets, or because they address directly other users. This group generally represents the common-user behavior ($x<1$), although information broadcasters ($x>1$) also exhibit the same \textit{low-activity} rate. 
These users fall in the lower region of the map.

Now that a reading of dynamic activity-connectivity maps has been provided, we can proceed with interpreting Fig.~\ref{fig:conn_activity}: the bottom-left quadrant reports the most common users, those with both activity and connectivity growth lesser than $1$. Conversely, the upper-right quadrant reports users with the higher connectivity growth and activity rates. These are influential ISIS supporters who are very active in the discussion. We note how the connectivity growth dimension spans three orders of magnitude in the positive domain, while the activity rate dimension only spans two orders of magnitude. This means that some users' followerships grows tens of times faster than the rate at which they follow others; conversely, the rate of receiving mentions is only up to tens of times higher than that of mentioning of others.
In the next section, we will devote special attention to these four different classes of users to determine what types of differences emerge in the ISIS social network.

\subsection{Dynamical Classes of ISIS Supporter Behaviors}\vspace{-3mm}

\begin{figure}[t]
\includegraphics[width=.49\columnwidth]{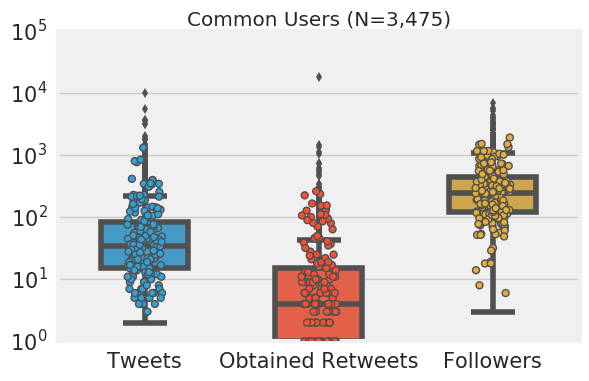}
\includegraphics[width=.49\columnwidth]{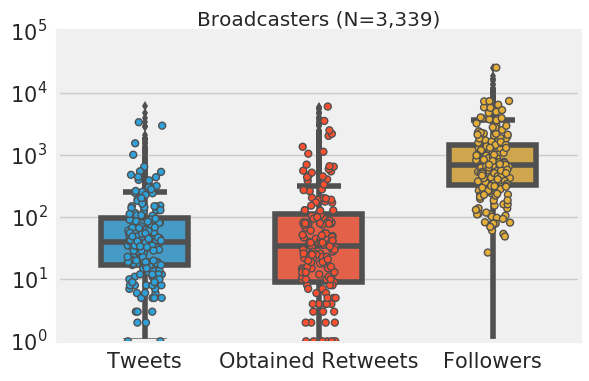}\\
\includegraphics[width=.49\columnwidth]{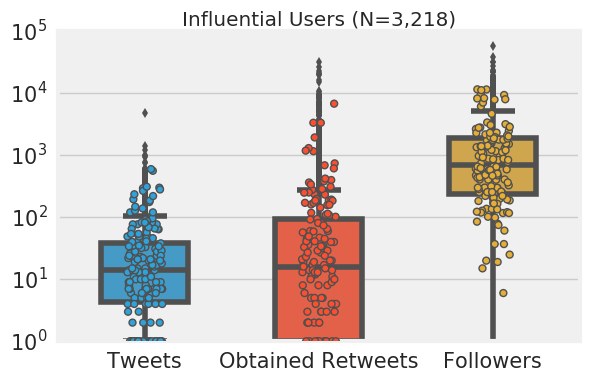}
\includegraphics[width=.49\columnwidth]{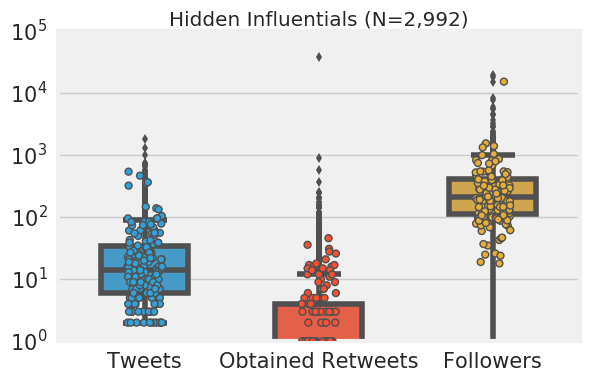}
\caption{Distributions of \textit{(i)} number of posted tweets, \textit{(ii)} number of obtained retweets, and \textit{(iii)} number of followers, for common users, broadcasters, influencial, and hidden influencial users, respectively.}\label{fig:boxplots}
\end{figure}

Prior research illuminated on the dynamical aspects of activity and connectivity in social media~\cite{lehmann2012dynamical, gonzalez2013broadcasters}. Next, we focus on the four classes of user behaviors highlighted by the dynamic activity-connectivity map. We first select, out of the twenty-five thousand ISIS supporters in our dataset, only the subset of those who have mentioned and have been mentioned at least once during the observation period. This will allow to focus on active accounts and correctly capture their activity rate. This filter reduces the number of users under investigation to $N=13,024$ ISIS supporters, nearly half of the entire ISIS population initially collected. We further divide these users in the four dynamical classes defined above. The classification is obtained by simply adopting the rules defining the four quadrants of Fig.~\ref{fig:conn_activity}, which yields $N=3,475$ common users ($x<1,y<1$), $N=3,339$ information broadcasters ($x>1,y<1$), $N=3,218$ influentials ($x>1,y>1$)), and $N=2,992$ hidden influential users ($x<1,y>1$). 

For each of these users, we generated the distribution of \textit{(i)} the total number of tweets they posted, \textit{(ii)} the cumulative number of times they have been retweeted, and \textit{(iii)} the maximum number of followers they gathered.
Fig.~\ref{fig:boxplots} shows the boxplots corresponding to the four dynamical classes.
Significant differences emerge: common users produce an amount of tweets very similar to that of broadcasters, but they accrue nearly one order of magnitude less retweets than the latter. Information broadcasters also appear to generate the largest followerships, on par with influential users; influentials, however, post significantly less tweets, while accumulating similar amounts of retweets than broadcasters, suggesting that our map successfully captures a notion of social influence intended as a proxy for attention generated to one's posts. The class of hidden influentials shows comparable activity to influential users, but significantly less influence, accruing about one order of magnitude less tweets and significantly smaller followerships than influentials.

Our analysis suggests that different classes of ISIS supporters' behaviors emerge. In the future, we will study what are the characteristics of different classes that produce the most effective propaganda and make the most influential users, analyzing content and language, political and religious beliefs, motives and attitudes of the ISIS social network.

\subsection{Adoption of ISIS Propaganda}\vspace{-3mm}

\begin{figure}[t] 
\includegraphics[width=\columnwidth]{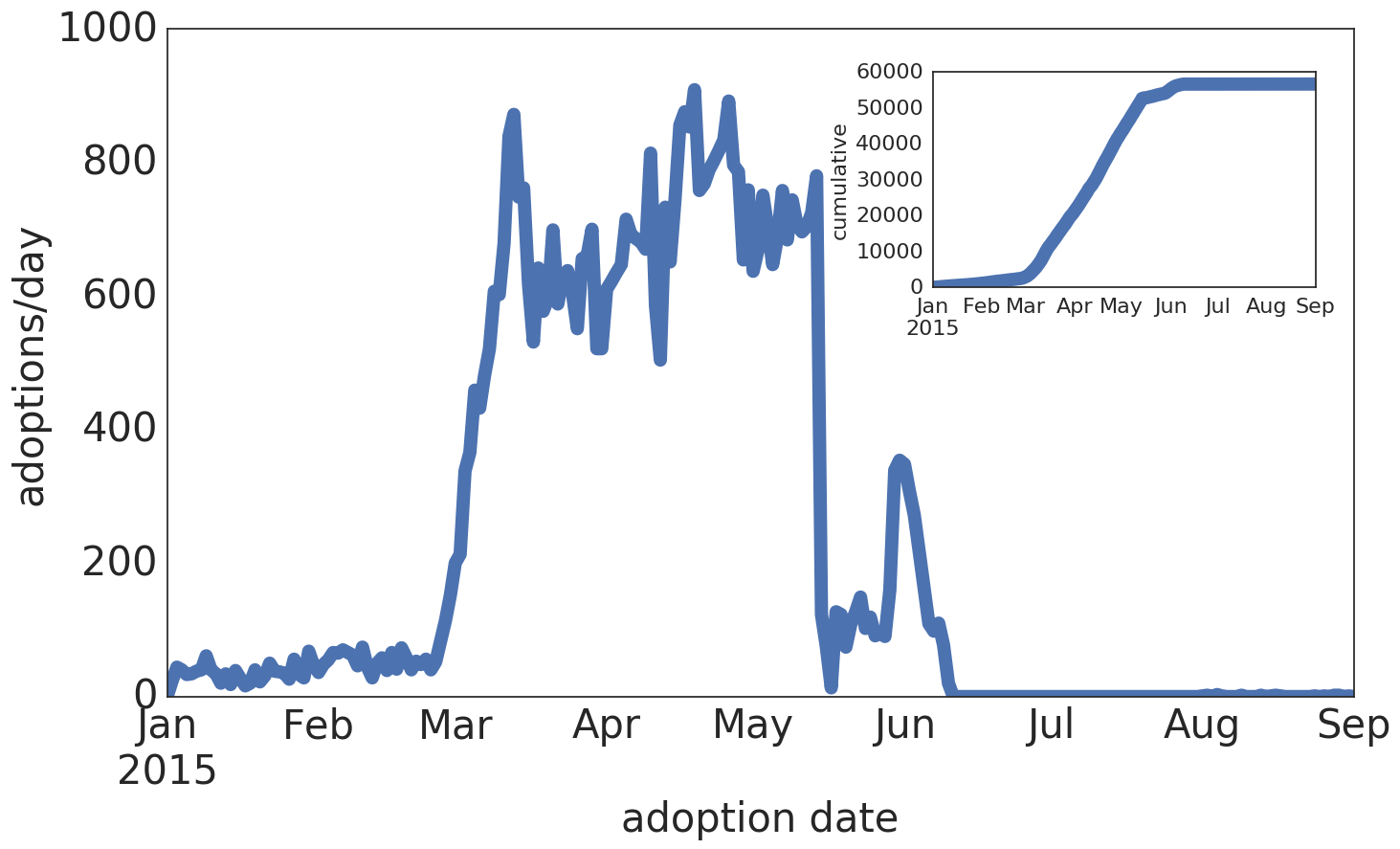}
\caption{Timeline of the number of individuals per day who adopted  ISIS contents.}
\label{fig:infected_timeline}
\end{figure}

So far, our analysis focused on characterizing some dimensions of the behavior of ISIS supporters on Twitter. 
Next, we investigate whether the content they generated has been adopted by other users who have become exposed to it. {\rev This will help address the last part of our second research question (\textbf{RQ2}), namely whether we can quantify the adoption of extremist content in the general population.}
Our notion of adoption is very simple: when a user who does not appear set of the twenty-five thousand ISIS accounts retweets for the first time any tweet generated by one such ISIS supporter, we count this as a content adoption. 
Although some recent work suggested that retweeting radical propaganda can be considered as an early sign of radicalization~\cite{berger2015isis, magdy2016failedrevolutions}, we call for caution about interpreting the results in such a way: this definition greatly simplifies the notion and complexity of such types of online adoption processes~\cite{centola2010spread, centola2011experimental, kramer2014experimental, ferrara2015measuring, ferrara2015quantifying}. However, we do believe that investigating the spread of radical content in the user population has a value to determine the extent and effectiveness of ISIS propaganda operations.

Fig.~\ref{fig:infected_timeline} shows the time series of the number of ISIS content adoptions per day during the first half of 2015. Prior to that, no significant amount of  adoptions could be observed, partly due to the low activity rate of the accounts under investigation.
In the three months between March and June 2015, we notice a significant uptake in the number of adoptions, peaking at nearly one thousand adoptions per day. In that period, at least 10,000 tweets per day (70,000-100,000 tweets/week) were generated by ISIS accounts (\textit{cf.} Fig~\ref{fig:total_tweets}). This suggests that a very significant fraction of tweets, about 5-10\%, was actually retweeted on average at least once by other users. During this period, a total of 54,358 distinct other users has retweeted at least once one of the twenty-five thousand ISIS supporters. 

If we simplify  propaganda diffusion as an infectious disease, we can draw a parallel with epidemics. 
The \textit{basic reproduction number} $R_0$ of an infection is the number of cases generated on average by an infected individual over the course of its infectious period, in an otherwise uninfected population. Given that 25,538 ISIS supporters generated 54,358 distinct \textit{infected users}, we can derive an $R_0 = 2.13$ for the ISIS propaganda ``infection''. In other words, an ISIS supporter before being suspended on average ``infected'' 2.13 other users. For comparison, health epidemics like Ebola,~\cite{althaus2014estimating, lau2017spatial} SARS,~\cite{wallinga2004different} HIV/AIDS,~\cite{hollingsworth2008hiv,prejean2011estimated} and certain strains of influenza~\cite{mills2004transmissibility} all have similar values of $2 < R_0 < 3$.

\begin{figure}[t] 
\includegraphics[width=\columnwidth]{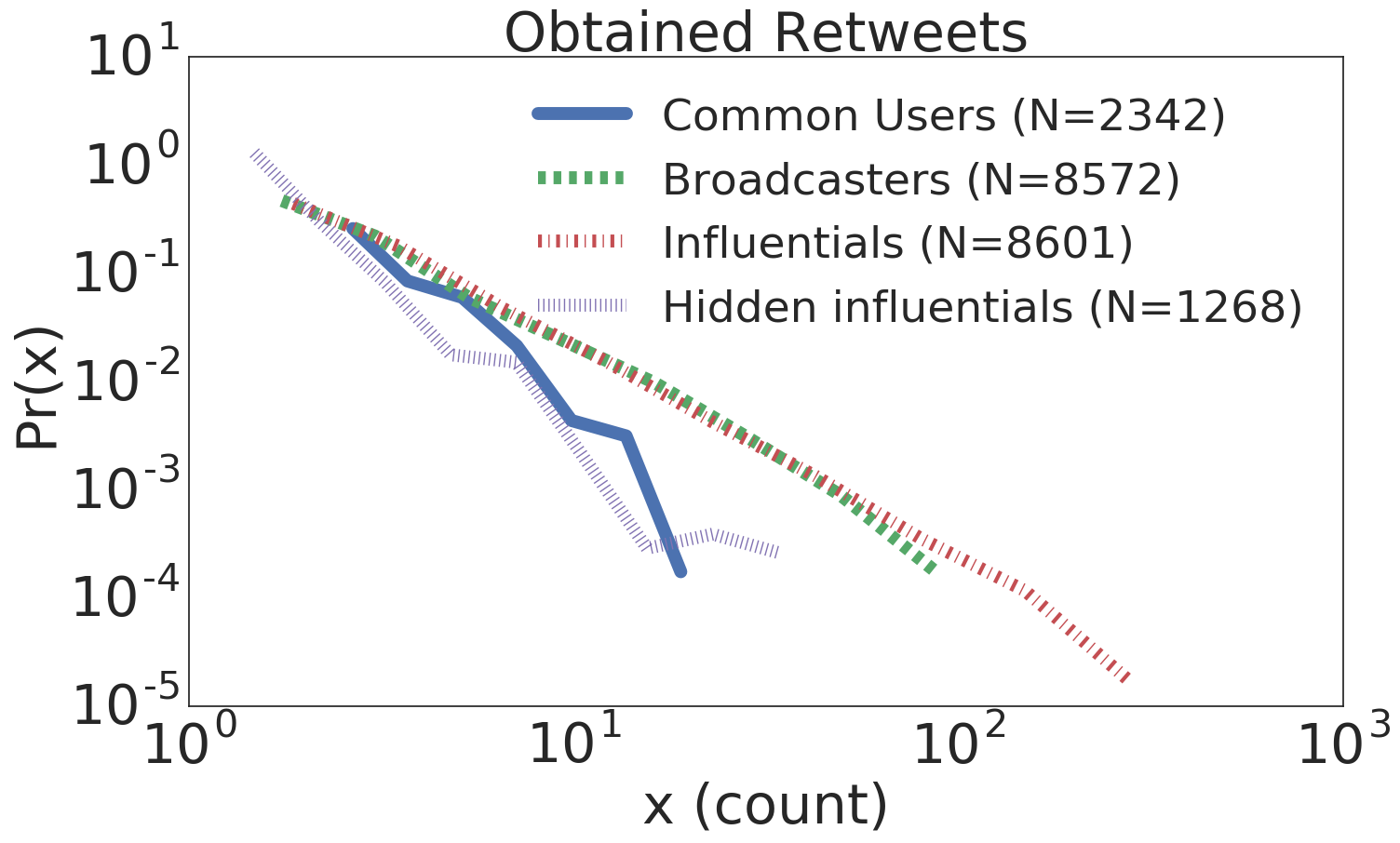}
\includegraphics[width=\columnwidth]{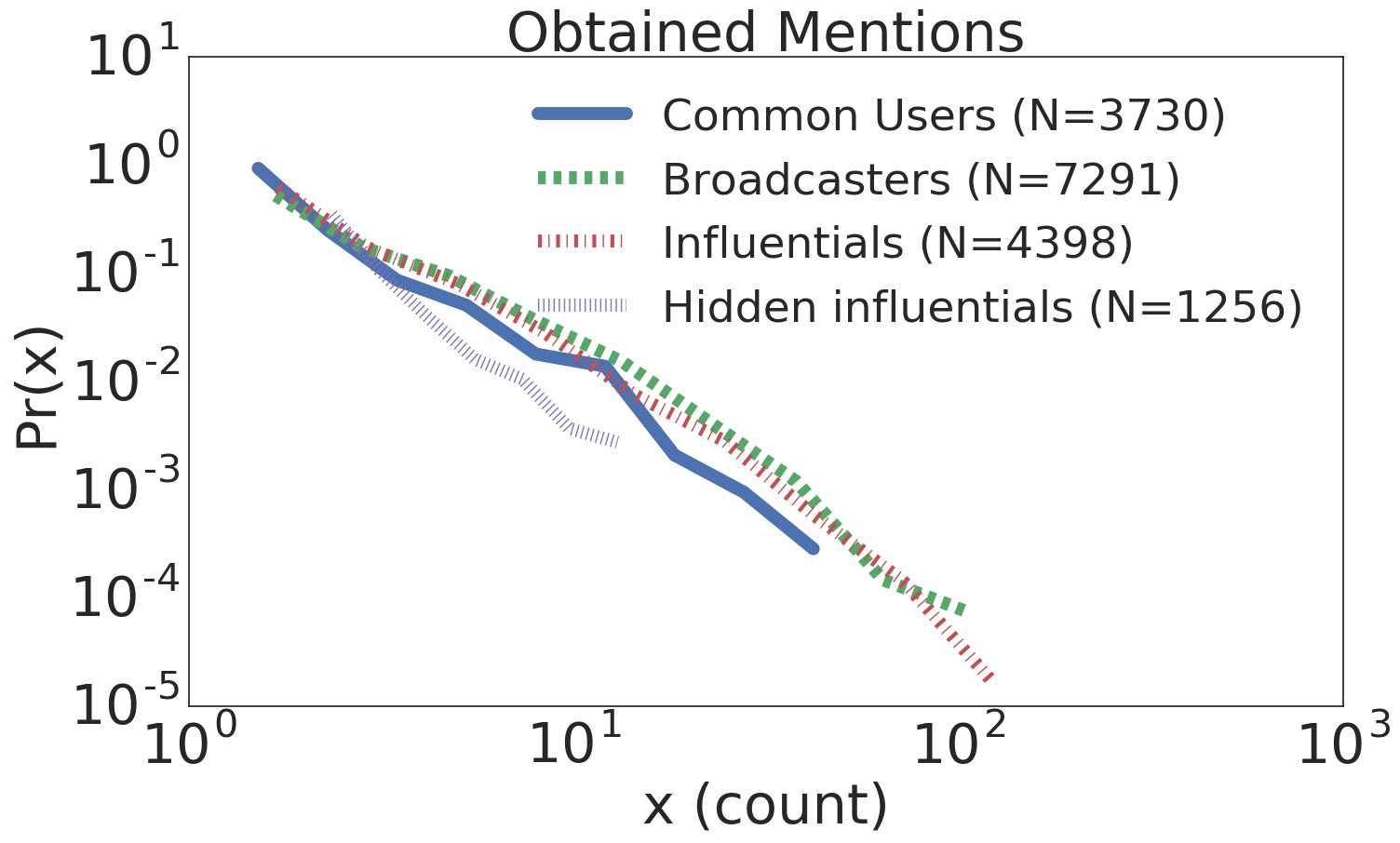}
\caption{\rev Probability distributions of retweets (top) and mentions (bottom) obtained by ISIS users as divided in the four dynamical classes, namely common users, broadcasters, influential and hidden influential users. Each group reports the size of the population under investigation (different in the two scenarios). }
\label{fig:contagion}
\end{figure}

\subsection{Contagion dynamics of ISIS Propaganda}
We conclude our analysis by studying the cascades of content adoptions generated by the four classes of users defined above. This investigation is twofold: first, we would like to determine whether the mechanisms of receiving retweets (what we defined as content adoption) and being mentioned by out-of-sample users exposed to extremist content follow the same or different dynamics. Second, we will compute the distributions of scores for the basic reproduction number $R_0$ relative to both receiving retweets and mentions for the four classes of users. The goal is to reveal whether any significant difference emerge between groups of users in their content spreading efficacy, and ultimately to understand which groups of users generated the most effective  information contagions.

Fig.~\ref{fig:contagion} shows the distributions of retweets (top) and mentions (bottom) received by the users in the four classes defined by means of the dynamic activity-connectivity map. Although overall all distributions are broad, as expected given the heterogeneous nature of information diffusion, the dynamics of obtaining retweets and mentions are significantly different for the four groups: in particular, for what concerns receiving mentions, no appreciable difference emerges among the four classes of users, which suggests that ISIS supporters are being mentioned in a similar fashion regardless of the class they belong to. However, receiving retweets shows a different mechanism: influential and broadcaster users receive generate significantly larger retweet cascades much more frequently than common users and hidden influentials.

\begin{figure}[t] 
\includegraphics[width=\columnwidth]{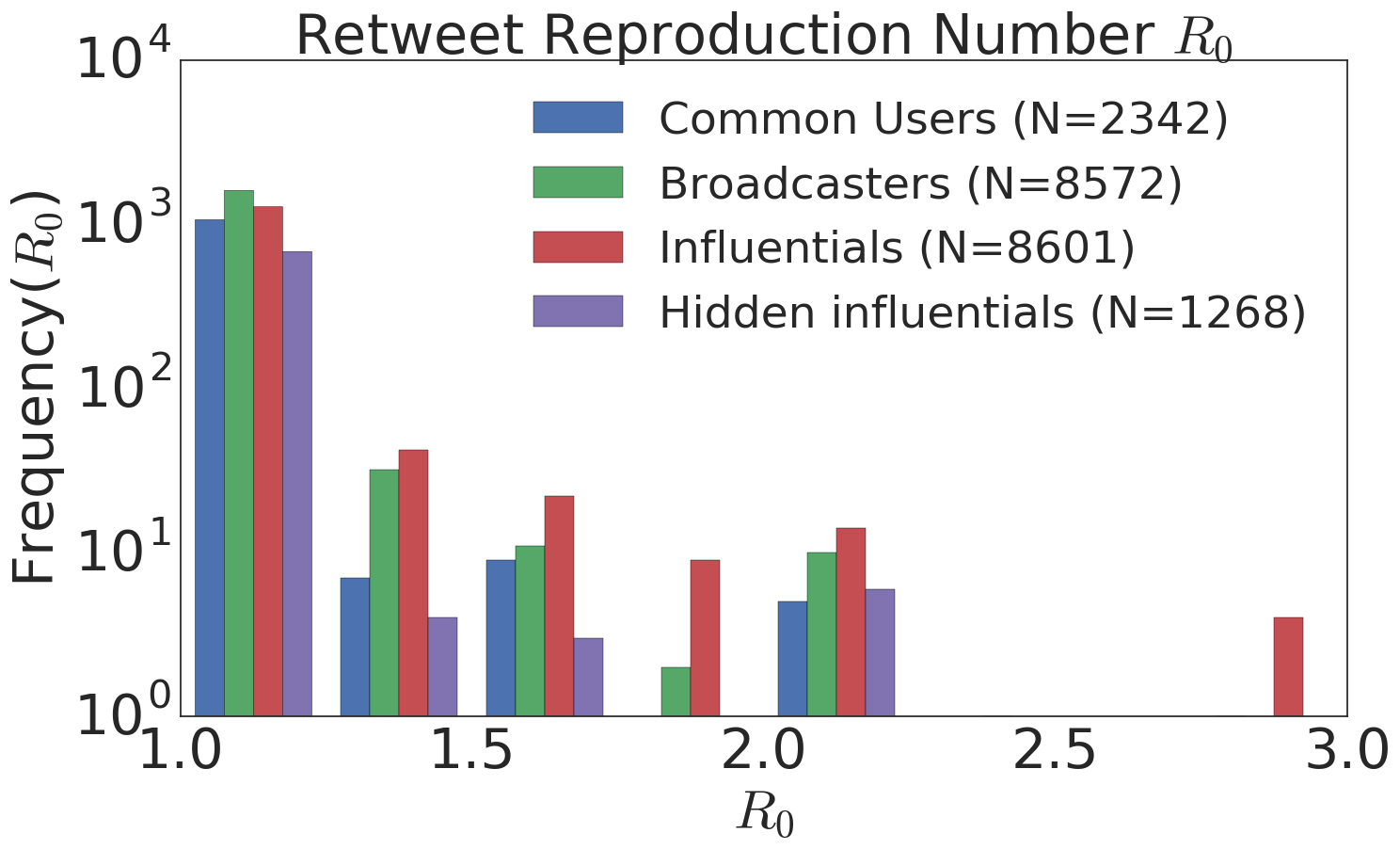}
\includegraphics[width=\columnwidth]{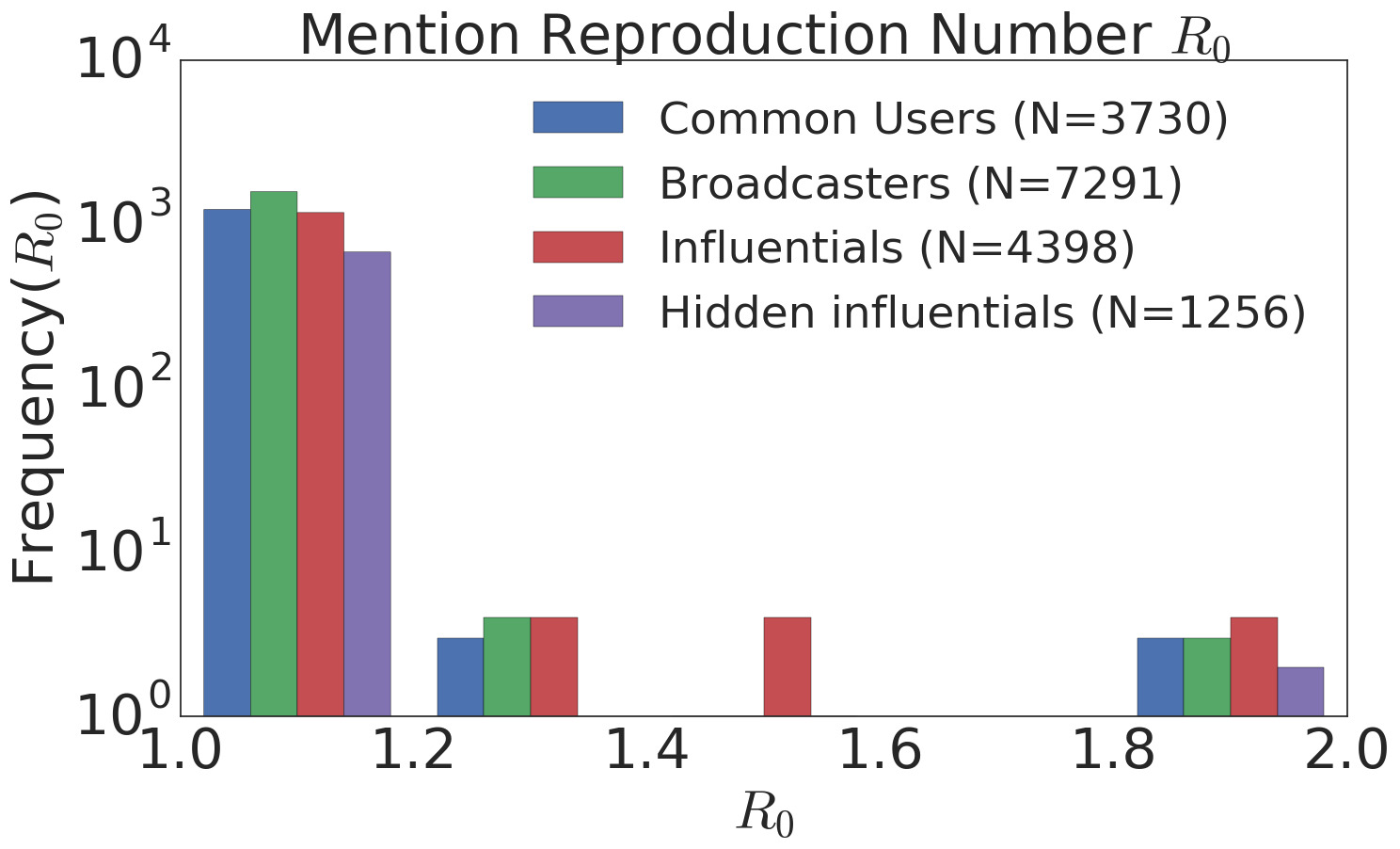}
\caption{\rev Frequency of basic reproduction scores $R_0$ relative to obtained retweets (top) and mentions (bottom) calculated for the four dynamical classes of ISIS users.}
\label{fig:contagion_r0}
\end{figure}

One question that thus rises concerns whether the actual contagion effectiveness varies between classes of users. In other words, what are the classes of users that are generating the most effective propaganda campaigns, in terms of adoption (\textit{i.e.,} retweets) and engagement (\textit{i.e.,} mentions)?
To this purpose, for each ISIS user who obtained at least one retweet (resp., mention), we calculated the fraction $\frac{RT}{T}$ of received retweets $RT$ (resp., mentions) over the total number of his/her tweets $T$ that have been retweeted at least once (resp., mentioned) by an out-of-sample user (\textit{i.e.,} a user not labeled as ISIS by our list). This measures the rate of diffusion of tweets in the otherwise uninfected population. We note that this is a simplification of the more traditional notion of information contagion where we would consider all the tweets generated and more complex diffusion mechanisms accounting e.g., for exposures, due to the limitation of the platform under study (namely, we do not have any information about information exposure on Twitter).
With some abuse of notation, we thus consider the fraction $\frac{RT}{T}$ to convey the meaning of the basic reproduction number $R_0$ typical of epidemiology.

{\rev Finally, ISIS accounts are divided in their four classes, according to the usual classification, and the frequency of the basic reproduction scores $R_0$ is shown in Fig.~\ref{fig:contagion_r0}, separately for the two dynamics of receiving retweets (top) and mentions (bottom)}. While no significant difference emerges, in either scenario, among the four different classes of users for the lowest scores (i.e., $2 < R_0 < 2.5$ for retweets, and $1 < R_0 < 1.2$ for mentions), strong class differences emerge for users whose content are more contagious ($R_0>2.5$ for retweets, and $R_0>1.2$ for mentions): concerning retweets, influential users, followed by information broadcasters, are receiving systematically more attention than users in other classes; the class differences for mentions are less pronounced.

\section{Related work}

One of the first computational frameworks, proposed by Bermingham \emph{et al.}~\cite{bermingham2009combining} in 2009, combined social network analysis with sentiment detection tools to study the agenda of a radical YouTube group: the authors examined the topics discussed within the group and their polarity, to model individuals' behavior and spot signs of extremism and intolerance, seemingly more prominent among female users. The detection of extremist content (on the Web) was also the focus of a 2010 work by Qi \emph{et al.}~\cite{qi2010hierarchical}. The authors applied hierarchical clustering to extremist Web pages to divide them into different pre-imposed categories (religious, anti immigration, etc.).

Scanlon and Gerber proposed the first method to detect cyber-recruitment efforts in 2014~\cite{scanlon2014automatic}. They exploited data retrieved from the Dark Web Portal Project~\cite{chen2008uncovering}, a repository of posts compiled from 28 different online fora on extremist religious discussions (e.g., Jihadist) translated from Arabic to English. After annotating a sample of posts as recruitment efforts or not, the authors use Bayesian criteria and a set of textual features to classify the rest of the corpus, obtaining good accuracy, and highlighted the most predictive terms.

Along the same trend, Agarwal and Sureka proposed different machine learning strategies~\cite{agarwal2014focused, sureka2014learning, agarwal2015using, agarwal2016spider} aimed at detecting radicalization efforts, cyber recruitment, hate promotion, and extremist support in a variety of online platforms, including YouTube, Twitter and Tumblr. Their frameworks leverage features of contents and metadata, and combinations of crawling and unsupervised clustering methods, to study the online activity of Jihadist groups on the platforms mentioned above.

A few studies explored unconventional data sources: one interesting example is the work by Vergani and Bliuc \cite{vergani2015evolution} that uses sentiment analysis (Linguistic Inquiry and Word Count \cite{tausczik2010psychological}) to investigate how language evolved across the first 11 Issues of Dabiq, the flagship ISIS propaganda magazine. Their analysis offers some insights about ISIS radicalization motives, emotions and concerns. For example, the authors found that ISIS has become increasingly concerned with females, reflecting their need to attract women to create their utopia society, not revolving around warriors but around families. ISIS also seems to have increased the use of internet jargon, possibly to connect with the identities of young individuals online.

Concluding, two very recent articles \cite{rowe2016mining, johnson2016new} explore the activity of ISIS on social media. The former~\cite{rowe2016mining} focuses on Twitter and aims at detecting users who exhibit signals of behavioral change in line with radicalization: the authors suggest that out of 154K users only about 700 show significant signs of possible radicalization, and that may be due to social homophily rather than the mere exposure to propaganda content. The latter study~\cite{johnson2016new} explores a set of 196 pro-ISIS aggregates operating on VKontakte (the most popular Russian online social network) and involving about 100K users, to study the dynamics of survival of such groups online: the authors suggest that the development of large and potentially influential pro-ISIS groups can be hindered by targeting and shutting down smaller ones.
For additional pointers we refer the interested reader to two recent literature reviews on this topic~\cite{correa2013solutions, agarwal2015applying}.

\section{Conclusions}
Since the appearance of the Islamic State (viz. ISIS), a consensus has emerged on the relationship between extremism and social media, namely that ISIS' success as a terrorist organization is due at least in part to its savvy use of social media. It is widely believed that ISIS has managed to increase its roster to tens of thousands of members by broadcasting its savage attacks over social media platforms such as Twitter, which helps radicalize and ultimately recruit fighters from around the world. 

Recent attacks on American and European soil demonstrate ISIS' potential to reach, organize, and mobilize lone wolves and sleeper terrorist cells among westerners. Many of these actors are known to have consumed radical material online and many have claimed to gravitate towards Islamic State because of it. 

{\rev This paper posed two research questions: the former was concerned with proposing good practices for data collection, validation, and analysis of online radicalization. The latter aimed at revealing the network and temporal activity patterns of ISIS influence on Twitter.
To address these questions we analyzed the activity of a group of twenty-five thousand users associated with ISIS}. These accounts have been manually identified, reported to Twitter for verification, and subsequently suspended due to their involvement with radical propaganda. This process yielded a human-curated dataset containing over three million tweets generated during a period of one and half year, 92\% of which in Arabic language.

{\rev Regarding the first research question (\textbf{RQ1}), we highlighted the challenges related to studying Arabic content, due to the limits of existing NLP and sentiment analysis toolkits. We therefore suggested the adoption of content-agnostic statistical and network techniques to dissect the users' temporal activities and connectivity patterns. By leveraging a computational tool named \textit{dynamic activity-connectivity map}, we highlighted the dynamics of social influence within ISIS support.  

For what concerns the second research question (\textbf{RQ2}),} our findings suggest complex strategies carried out by these users to manipulate and influence others: four dynamical classes of ISIS supporters emerged (common sympathizers, information broadcasters, influential and hidden influential users), each with distinct activity and connectivity  patterns. We concluded by quantifying the extent to which ISIS support and extremist content are adopted in the general population: by drawing a parallel between propaganda and epidemics spreading, we determined that each ISIS supporter ``infected'' on average 2.13 other users before Twitter suspended his/her account, highlighting that receiving retweets and mentions follow different dynamics, and that broadcasters and influential users generate much more widespread contagions.

Although we call for caution in the interpretation of these results, due to the great simplifications introduced by our framework, we believe that our findings will help design and implement effective countermeasures and responses to ISIS social media operations and other forms of online extremist propaganda.

\newpage
\section*{Acknowledgments} 
The author is grateful to Max Abrahms (Northeastern University) for useful discussions, and to Alessandro Flammini and Onur Varol (Indiana University) for their support in collecting the Twitter dataset. 
This work has been partly funded by the Office of Naval Research (ONR), grant no. N15A-020-0053. 
This research is also based upon work supported in part by the Office of the Director of National Intelligence (ODNI), Intelligence Advanced Research Projects Activity (IARPA). The views and conclusions contained herein are those of the authors and should not be interpreted as necessarily representing the official policies, either expressed or implied, of ODNI, IARPA, ONR, or the U.S. Government. The U.S. Government had no role in study design, data collection and analysis, decision to publish, or preparation of the manuscript. The U.S. Government is authorized to reproduce and distribute reprints for governmental purposes notwithstanding any copyright annotation therein. 

\bibliographystyle{abbrv}
\bibliography{data_science}

\end{document}